\documentclass[twoside]{article}
\usepackage{aistats2018}
\usepackage{amsmath, amssymb, amsfonts}
\usepackage{graphicx}
\usepackage[style=nature,backend=bibtex,natbib=true,sorting=nty]{biblatex} 
\bibliography{Bibliography.bib}
\defbibheading{bibliography}[\refname]{}

\begin{document}

%

%

\twocolumn[

\aistatstitle{Integrative analysis of time course metabolic data and biomarker discovery} 

\aistatsauthor{ Takoua Jendoubi \And Timothy M.D. Ebbels }

\aistatsaddress{ School of public health\\
Imperial College London \And  Departement of surgery and cancer \\Imperial College London } 
]

\begin{abstract}
  Metabolomics time-course experiments provide the opportunity to understand the changes to an organism by observing the evolution of metabolic profiles in response to internal or external stimuli. Along with other omic longitudinal profiling technologies, these techniques have great potential to complement the analysis of complex relations between variations across diverse omic variables and provide unique insights into the underlying biology of the system. However, many  statistical methods currently used to analyse short time-series omic data are i) prone to overfitting or ii) do not take into account the experimental design or iii) do not make full use of the multivariate information intrinsic to the data or iv) unable to uncover multiple associations between different omic data. The model we propose is an attempt to i) overcome overfitting by using a weakly informative Bayesian model, ii) capture experimental design conditions through a mixed-effects model, iii) model interdependencies between variables by augmenting the mixed-effects model with a conditional auto-regressive (CAR) component and iv) identify potential associations between heterogeneous omic variables by using a horseshoe prior.\\
\textbf{Results:} We assess the performance of our model on synthetic and real datasets and show that it outperforms existing models for metabolomic longitudinal data analysis. Our proposed method is able to identify metabolic biomarkers related to treatment, infer perturbed pathways as a result of treatment and find significant associations with additional omic variables. We also show through simulation that an informative beta like prior compares better than a non-informative uniform prior in inferring significant pathways. On real data, we investigate how the number of profiled metabolites can affect the predictive ability of the model.\\
Supplementary material for this article are available online.
\end{abstract}
\section{INTRODUCTION}
Over the past years, there has been a significant development in high-throughput omics technologies e.g. metabolomics, transcriptomics, genomics, epigenomics and proteomics along with a growing interest into joint modelling of multi-omic data \citep{joyce2006,ebrahim2016}. In metabolomics, several approaches are used to understand the response of a biological system as a function of an internal or external perturbation by monitoring ``the chemical fingerprints that specific cellular processes leave behind" \citep{daviss2005}. These chemical fingerprints are most commonly interrogated in terms of metabolite (i.e. low weight molecules) concentration, structure and transformation pathways (i.e set of chemical reactions) in order to identify a biomarker related to the studied process. Biomarker discovery consists of identifying a metabolite that has significant association patterns with a particular phenotype (disease, clinical variables, physical traits, etc) and that can be thus used as an indicator of that specific phenotype. Typical experimental platforms use analytical techniques such as nuclear magnetic resonance spectroscopy (NMR) \citep{reo2002} and mass spectrometry (MS) \citep{dettmer2007} to generate appropriate spectral metabolomic profiles of the studied biological system. 

Metabolomic data sets are characterized by high correlation structures in that many spectral peaks can arise from the same metabolite and metabolites operate within networks of chemical reactions. In addition, further correlation structure is present in longitudinal metabolomic studies due to repeated measurements of observations over time. Additional challenges include not only the low number of time points and samples compared to the number of profiled metabolic variables, but also integration of a different omic data to the metabolomic data. 

First, metabolomic time series are often short due to experimental costs or ethical considerations. Typically, less than 10 time points are available compared to a large number of metabolic variables profiled at each time point e.g. hundreds of metabolic variables for targeted experiments and thousands of metabolic variables for untargeted experiments. Taking into account the small number of time points compared to the large number of metabolic variables profiled, the number of temporal patterns that will arise is limited (due to the limited number of degrees of freedom). Some temporal patterns will be repeated and thus these patterns can be induced by randomness. Second, models fitted to a small number of data points are often prone to overfitting i.e. the model is very sensitive to small fluctuations. This can lead to a poor fitting to unseen data and a high generalisation error. Third, it is also important to consider the number of parameters of the statistical model and make use of the simplest models in order to avoid overparametrisation. 
Finally, monitoring heterogeneous omic variables can substantially enhance the understanding of the underlying biological mechanism and provide a systems biology approach as these omic variables represent entities that are often involved in related cellular processes \citep{joyce2006, ebrahim2016}.

For all these reasons, metabolomics scientists need cautious estimates model parameters in order to ensure robust interpretation of the results. Hence, appropriate models are needed to integrate heterogeneous omic data and take into account both experimental conditions and biological variations in order to extract important information from the data. In this paper, we are interested in dose-response time course experiments where additional omic variables (bacteria, genes, transcripts, etc) are monitored along with metabolites in the context of biomarker discovery.\\
 The main contribution of our work is a \textit{single} probabilistic generative model that i) can overcome overfitting via the use of weakly informative priors ii) makes use of mixed effects models to model the experimental design iii) models metabolite interactions by using pathway information through a conditional auto-regressive (CAR) component and iv) uncovers multiple associations between metabolites and other omic variables by using a horseshoe prior.  An additional benefit of our approach is that it naturally yields a list of perturbed metabolic pathways since it is based both on a mixed effects component and a CAR component.


\section{RELATED WORK}
There is a growing interest in longitudinal experiments for heterogeneous omics data and statistical models to infer biomarkers of a particular treatment or disease over time. Different approaches attempt to infer influential or significant metabolites using dynamic metabolomic data under the assumption that metabolites are independant. These models include fitting smooth splines mixed effects models (SME) to time curves \citep{berk2011} and linear mixed effects models augmented with a variable selection approach \citep{Mei2009} . However between-metabolite correlation is richly structured and biologically relevant and should be modelled. 

Seemingly unrelated regression accounts for metabolite correlation by using correlated regression errors and can be used to identify biologically significant metabolites \citep{Chen2015, Chen2017}. In gene expression data analysis, \cite{Pham2015} recently proposed to use confirmatory factor analysis to capture gene-pathway relationship and a conditional autoregressive model to capture relationships between a set of pathways where pathway network has been constructed based on KEGG \citep{kanehisa2000} pathways. The latter accounts for biological variation in the data and allows ease of interpretation.

 In the metabolomics literature, traditional frameworks for metabolomic data analysis use dimensionality reduction techniques, namely principal component analysis (PCA), partial least squares (PLS) \citep{wold1983} and PLS derived models (OPLS \citep{trygg2002}, O2PLS \citep{trygg2003}, OnPLS \citep{lofstedt2011}) to take into account high correlation between metabolites. Extension of PLS to O2PLS and OnPLS allows for integrative analysis of heterogeneous omic data. PCA and PLS models are very popular among metabolomicists as tools for exploratory data analysis, visualisation purposes and ease of interpretation. One of the interests of PCA (PLS) derived models is to be able to visually assess whether or not there is a time effect in the data and identify metabolites that change over time by looking into time trajectories of each metabolite \citep{antti2002}.
 
 Extensions to PCA (PLS) for longitudinal analysis include lagged PCA (PLS) and dynamic PCA (PLS) where a backshift matrix is introduced to take into account time dependancy. Similarly, \cite{tim2008} used a set piecewise orthogonal projections latent structures to describe changes between neighboring time points. PARAFAC \citep{bro1997} is a multi-linear unsupervised decomposition method that can account for the multi-way variation seen in dynamic metabolomics data. Recently, dynamic probabilistic PCA (DPPCA) was proposed in \cite{2014dynamic} as a generative probabilistic model of longitudinal metabolomic data where a stochastic volatility model is used for the latent variables. The main inconvenience of these approaches is that further techniques such as multiple testing correction have to be separately applied to the data in order to identify biomarkers and they do not take heterogeneous data (i.e. data from different omics techniques) into account.

 In contrast, we here provide a \textit{single} model that takes into account metabolite interactions, time variation and experimental design, infers influential metabolites and also quantifies relationships between metabolites and additional  omic variables (if any).

\section{MODEL FRAMEWORK}
Given a metabolomics data $\boldsymbol{X} \in \mathbb{R}^{N \times T \times M}$ where $N$ is the number of observations, $T$ the number of time points and $M$ the number of metabolites. $\boldsymbol{Y} \in \mathbb{R}^{N \times T \times K}$ is an additional continuous omic data measured along with $\boldsymbol{X}$ where $K$ is the number of associated omic variables. The set of $N$ observations consists of a set of cases and controls. Throughout the paper, index $i$ always runs through observations, index $t$ runs through time points, index $m$ runs through metabolites and index $k$ runs through $\boldsymbol{Y}$ variables. Vector quantities are written in bold. Matrices are written in bold capitals.
Our goal is to build a simple model that can identify biomarkers relative to a specific treatment in time taking into account the multiple sources of variations in the data.

The model is built on three levels: First, a CAR component to capture interaction between metabolites. Second, a variable selection model to uncover associations between metabolites and $\boldsymbol{Y}$ data. Third, a mixed effects component to model experimental design. We give more details about each level of our model respectively in each of the following sections.

%

\subsection{Metabolite interactions}
In a similar fashion to \cite{Pham2015}, we model metabolite interactions via the CAR model. In fact, incorporation of pathway information in the CAR model through the variance matrix helps provide chemists with an easily interpretable model. First, we assume that the concentration of each metabolite is linearly influenced by concentration levels of metabolites in the same pathway. Let $\boldsymbol{C} \in \mathbb{R}^{M \times M}$ be the design matrix quantifying metabolite interactions such that matrix elements $c_{mm} = 0$, $c_{mj} \neq 0$ if metabolites $m$ and $j$ are in the same pathway and $0$ otherwise. Thus, metabolite intensity levels can be expressed as:
\begin{equation}
x_{itm} \vert  \boldsymbol{x}_{it,-m}, \boldsymbol{\mu}_{it}, \boldsymbol{C}, \sigma  \sim  N(\mu_{itm} + \sum_{\substack{j=1\\ j \neq m}}^M c_{mj} (x_{itj}-\mu_{itj}), \sigma^2)
\end{equation}
where $\boldsymbol{x}_{it,-m}$ represents measurements of metabolites of sample $i$ at time point $t$ excluding metabolite $m$, $\mu_{itm}$ is a function of covariates of sample $i$ for metabolite $m$ at time point $t$. If we define $\boldsymbol{I}_M$ the $M$th order identity matrix, the joint distribution of $\boldsymbol{x}_{it}$ can be explicitely written as \citep{cressie2015}:
\begin{eqnarray}
\boldsymbol{x}_{it} \vert \boldsymbol{\mu}_{it}, \boldsymbol{C}, \sigma  \sim  N \left(\boldsymbol{\mu}_{it} , \left( \boldsymbol{I}_M-\boldsymbol{C} \right)^{-1} \sigma^2 \right)
\end{eqnarray}
Chemists are most interested by identifying which pathways are ``on" or ``off" as an effect of treatment. In the CAR literature, the design matrix $\boldsymbol{C}$ can be modeled as a scaled 
product of a diagonal weight matrix and an
adjacency matrix. In order to allow for pathway perturbation inference we construct the distance matrix based on the individual contribution of each pathway. To be precise, we define 
$\boldsymbol{C} \left( \boldsymbol{\phi} \right) = \sum_{p=1}^{P}\phi_p \boldsymbol{G_p}\boldsymbol{A_p}$ where $P$ is the number of pathways. 
The distance matrices $\boldsymbol{A_p}$ are a zero-diagonal symmetric adjacency matrices with elements $a_{mj}^p$ equal to the inverse of the length of the shortest path between metabolites $m$ and $j$ if they are in pathway $p$ and $0$ otherwise. A path between two metabolites consists in the number of reactions that lead from one metabolite to the other. The shortest path is the path that contains the smallest number of reactions. In the diagonal matrices $G_p$ we use the reciprocal of the number of neighbors of each metabolite in pathway $p$ i.e  $ \left(g_{mm}^{p} \right)^{-1}= \sum_{j=1}^M (a_{mj}>0) $ so that the squared partial correlation $\text{cor} \left( x_{itm}, x_{itj} \vert \boldsymbol{x}_{it,-(m,j)} \right) ^2 \propto \phi_p^2 g_{mm}^p g_{jj}^p $ is reduced when more metabolites from the same pathway are profiled \citep{cressie2015}. 
The model parameter
$\boldsymbol{\phi} = \lbrace \phi_p \rbrace_{p=1}^P$ is estimated. It quantifies pathway contribution and is referred to as \textit{spatial}-dependence parameter in the CAR literature.

 The model needs to comply with the condition that $\boldsymbol{I}_M-\boldsymbol{C\left( \boldsymbol{\phi}\right)}$ is positive definite. If we assume that pathways are \textit{a priori} equally perturbed, $\phi_p$ must fall in the interval $\left( \frac{1}{P\xi_p^1}, \frac{1}{P\xi_p^2} \right)$ where $\xi_p^1$ and $\xi_p^2$  are the minimum and maximum eigenvalues of $\boldsymbol{G_p} \boldsymbol{A_p}$, respectively.  In practice, strong interaction between observed metabolites of pathway $p$ is reproduced in CAR models only when the scaling parameter $\phi_p$ is quite close to one of the boundaries $\frac{1}{P\xi_p^1}, \frac{1}{P\xi_p^2}$. Hence, we use a beta-type prior for $\phi_p$ that places substantial mass on large values of $\vert \phi_p \vert$ \citep{banerjee2014} :
\begin{eqnarray}
\text{p} \left( \phi_p \right) = \dfrac{1}{\textbf{B} \left(\frac{1}{2}, \frac{1}{2} \right)} \left( \phi_p- \frac{1}{P\xi_p^1} \right)^{-\frac{1}{2}} \left(  \frac{1}{P\xi_p^2} - \phi_p \right)^{-\frac{1}{2}} 
\label{eqphip}
\end{eqnarray}
where \textbf{B} is the beta function. The parameter $\sigma^2$ captures variance heterogeneity in metabolite intensities and is given an inverse gamma prior $ \textbf{G} \left( \psi, \psi-1 \right)$. This prior provides $2\psi$ pseudo-observations in addition to $NT$ available observations. In order to build a reasonably informative prior we set $\psi = N \times T/4$.

\begin{figure}[!t]
\includegraphics[scale=0.2]{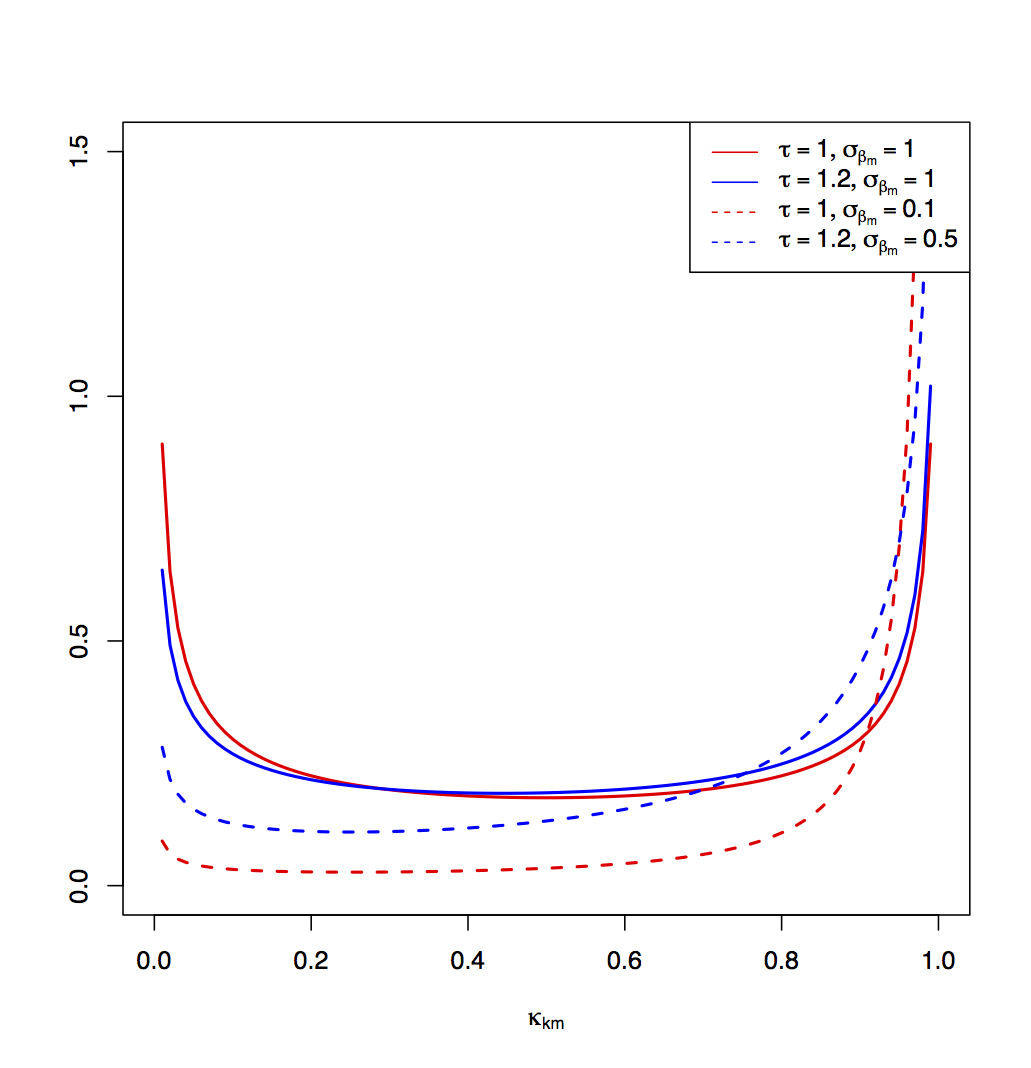}
\caption{Prior $\text{p} \left( \kappa_{mk} \vert \tau, \sigma_{\beta_m} \right)$ on $\kappa_mk$ for different values of $\sigma_{\beta_m}$ and $\tau$.  The prior distribution skews towards 1 if $\tau$ increases or $\sigma_{\beta_m}$ decreases.}
\label{fig1}
\end{figure}

\begin{figure}[!t]
\centering
\includegraphics[scale=0.2]{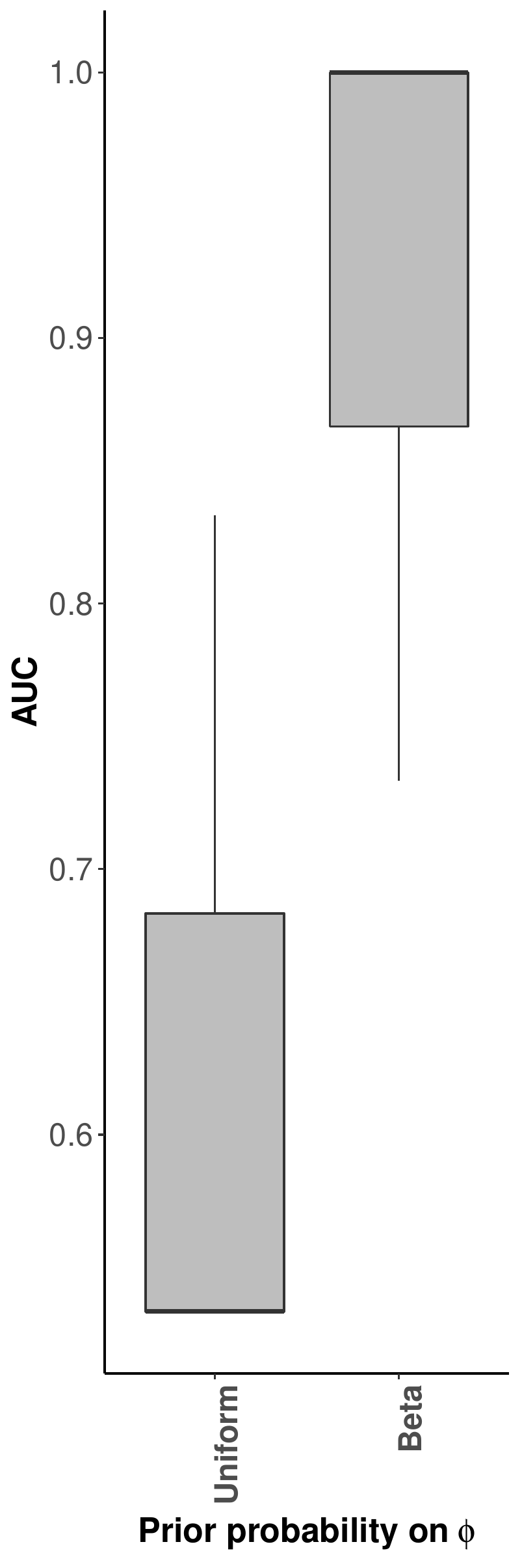}
\includegraphics[scale=0.27]{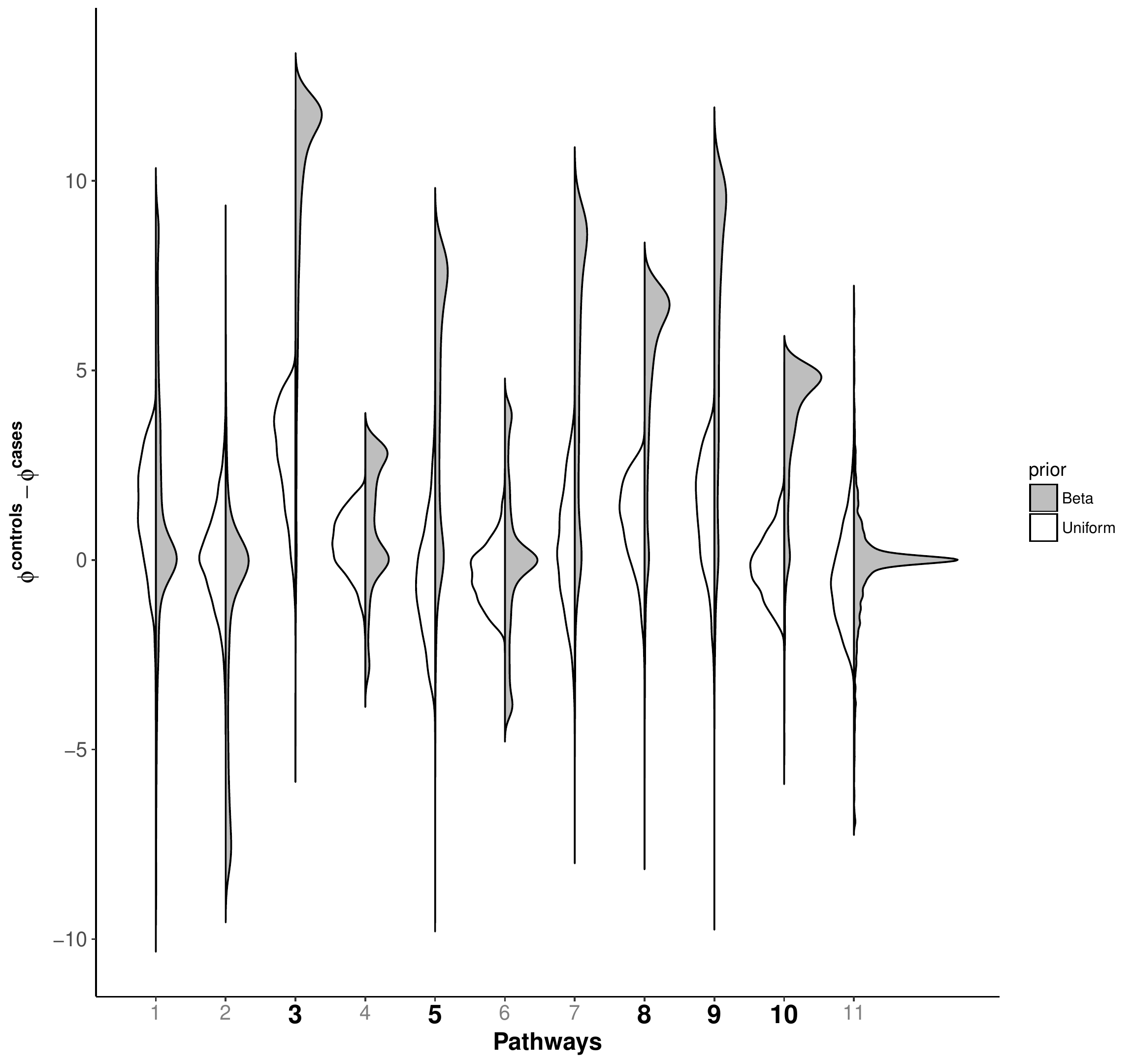}
\caption{Left : Boxplots of Area Under the Curve (AUC) for pathway perturbation inference with uniform prior on $\boldsymbol{\phi}$ compared to a beta like prior on $\boldsymbol{\phi}$ for 10 simulated datasets. We infer perturbations based on the posterior distribution of $ \boldsymbol{\phi}^{\text{controls}} - \boldsymbol{\phi}^{\text{cases}}$. Right : Distribution of $ \boldsymbol{\phi}^{\text{controls}} - \boldsymbol{\phi}^{\text{cases}}$ under the uniform prior (white) and the beta like prior (grey). True perturbed pathways are printed in bold on the x axis.}
\label{fig2}
\end{figure}

\subsection{Integrative analysis}
Interactions between heterogeneous omic variables such as transcripts and metabolites or gene expression and metabolites is modeled via the following hierarchical shrinkage model:
\begin{eqnarray}
\mu_{itm} = \alpha_m + \gamma_{im} + \boldsymbol{y}_{it}\boldsymbol{\beta}_m + \nu_{itm}\\
\label{eq8}
\beta_{mk} \vert \lambda_{mk}, \sigma_{\beta_m} \sim N \left( 0, \lambda_{mk}^2 \sigma^2_{\beta_m} \right)\\
\lambda_{mk} \vert \tau \sim \text{St}^+ \left( \tau ,0,1 \right)
\end{eqnarray}

where $\text{St}^+$ denotes the half Student-t distribution with $\tau$ degrees of freedom, $\alpha_m$ represents treatment effect for metabolite $m$, $ \gamma_{im} \sim N\left(0, \sigma_{\gamma_m}^2 \right)$ represents individual perturbations for metabolite $m$, $\nu_{itm} \vert  \nu_{i,t-1,m}  \sim  N\left(\theta_m \nu_{i,t-1,m}, \sigma_{\nu_m}^2\right)$ follows and auto-regressive process and represents temporal effects for metabolite $m$ of individual $i$ at time point $t$. Finally, $\boldsymbol{\beta}_m$ quantifies interactions between metabolite $m$ and other omic variables. $\lambda_{mk}$ is called local shrinkage parameter whilst $\sigma^2_{\beta_m}$ is the global shrinkage parameter. For $\tau=1$, this prior reduces to the horseshoe prior. Intuitively, for small values of $\lambda_{mk}$ the coefficient $\beta_{mk}$ is very close to $0$ while for relevant variables $\lambda_{mk}$ will be large. In addition, $\sigma_{\beta_m}$ controls the overall shrinkage level i.e sparsity of the vector $\boldsymbol{\beta}_m$ is more important for small values of $\sigma_{\beta_m}$.

Define $\kappa_{mk} = \dfrac{1}{1+ \lambda_{mk}^2 \sigma^2_{\beta_m}/ \tau}$ a random shrinkage coefficient such that $\kappa_{km}\approx 0$ when $\lambda_{mk}$ is large and $\kappa_{km}\approx 1$ when $\lambda_{mk}$ is small. This transformation implies the following prior distribution on $\kappa_{mk}$:
\begin{equation}
\text{p} \left( \kappa_{mk} \vert \tau, \sigma_{\beta_m} \right) = \dfrac{1}{2 \sqrt{\pi} \textbf{B} \left( \frac{\tau}{2}, \frac{1}{2} \right)} \dfrac{\sigma_{\beta_m}^{\tau} \kappa_{mk}^{\tau/2-1} \left( 1-\kappa_{mk}\right)^{-1/2}}{\left( 1- \kappa_{mk} + \kappa_{mk} \sigma_{\beta_m}^2 \right)}
\end{equation}
This prior density is shown in figure \ref{fig1} for different values of $\sigma_{\beta_m}$ and $\tau$. It reduces to a Beta$\left( \tau/2, 1/2 \right)$ distribution if $\sigma_{\beta_m}=1$ and to a Beta$\left( 1/2, 1/2 \right)$ which looks like a horseshoe, if in addition $\tau=1$. When $\tau$ increases, Beta$\left( \tau/2, 1/2 \right)$ skews towards $1$ which increases the global shrinkage power. The expectation of $ \boldsymbol{\beta}_m $ given $\boldsymbol{Y}, \boldsymbol{\kappa}_m, \tau, \boldsymbol{\mu}_{tm} $ can be expressed as:
\begin{eqnarray}
\mathbb{E} \left( \boldsymbol{\beta}_m \vert \boldsymbol{Y}, \boldsymbol{\kappa}_m, \tau, \boldsymbol{\mu}_{tm} \right) &=& \left( \sum_{t=1}^T \boldsymbol{Y}_t^T \Sigma_m^{-1} \boldsymbol{Y}_t + \tau \Upsilon_m  \right)^{-1} \nonumber \\
& & \times \sum_{t=1}^T \boldsymbol{Y}_t^T  \Sigma_m^{-1} \boldsymbol{\mu}_{tm}
\label{eq12}
\end{eqnarray}

where $\Sigma_m = \left( \frac{\sigma_{\nu_m}^2}{1-\theta_m^2} + \sigma_{\gamma_m}^2 \right) \boldsymbol{I}_N$ and $\Upsilon_m$ is a diagonal matrix of order $K$ with elements $1/\kappa_{mk}-1$. Equation~(\ref{eq12}) introduces a penalty term $ \tau \Upsilon_m$ where $\Upsilon_m$ is a metabolite specific penalty term introduced by the horseshoe prior and $\tau$ is a global penalty term. Precisely, $\tau$ captures the overall sparsity level amongst all metabolites.  The expectation of $ \boldsymbol{\beta}_m $ given $\boldsymbol{Y}, \boldsymbol{\kappa}_m, \tau, \boldsymbol{\mu}_{tm} $ is very similar to the estimate of $ \boldsymbol{\beta}_m $ under ridge regression where  $ \tau \Upsilon_m$ simply reduces to $\tau \boldsymbol{I}_N$.\\
The global sparsity level can be controlled using $\tau$. Increasing the global sparsity level is a desired property in omic studies, as usually we deal with a large number of omic variables where only few are important.  
Moreover, when there is prior knowledge available, specifying $\tau$ \textit{a priori} can optimize the inference and additionally, provide a more informative prior on $\lambda_{mk}$. If we fix $\text{p}\left( \sigma_{\beta_m}^2\right) \propto 1/\sigma_{\beta_m}^2$, integrating over $\sigma_{\beta_m}$ gives the expected value of $\kappa_{mk}$ as :
\begin{eqnarray*}
\mathbb{E} \left( \kappa_{km} \vert \tau \right) = \frac{\Gamma \left(1/2\right)^{-1}}{2 \sqrt{\pi} \Gamma \left(\tau/2\right)} \\ \times \textbf{G}_{3,3}^{2,3} 
\left(
\begin{matrix}
1, \tau/2, 0 \hfill \\ \tau/2, \tau/2-1/2, 0
\end{matrix}
\, \middle\vert \,
1-\sigma_{\beta_m}^2
\right) 
\label{eq9}
\end{eqnarray*}
where $\textbf{G}_{\cdot,\cdot}^{\cdot,\cdot} $ is Meijer's G-function \citep{meijer1936}.
The equation above 
can be used to fix $\tau$ \textit{a priori} by defining the expected proportion of shrunk coefficients. In practice, different values of $\tau$ are plugged into the equation above 
to get the desired proportion of shrunk coefficients. However, many definite integrals can be obtained using the tables of Meijer functions in \cite{brychkov2008} for special values of parameters.

\begin{figure}[!t]
\centering
\includegraphics[scale=0.3]{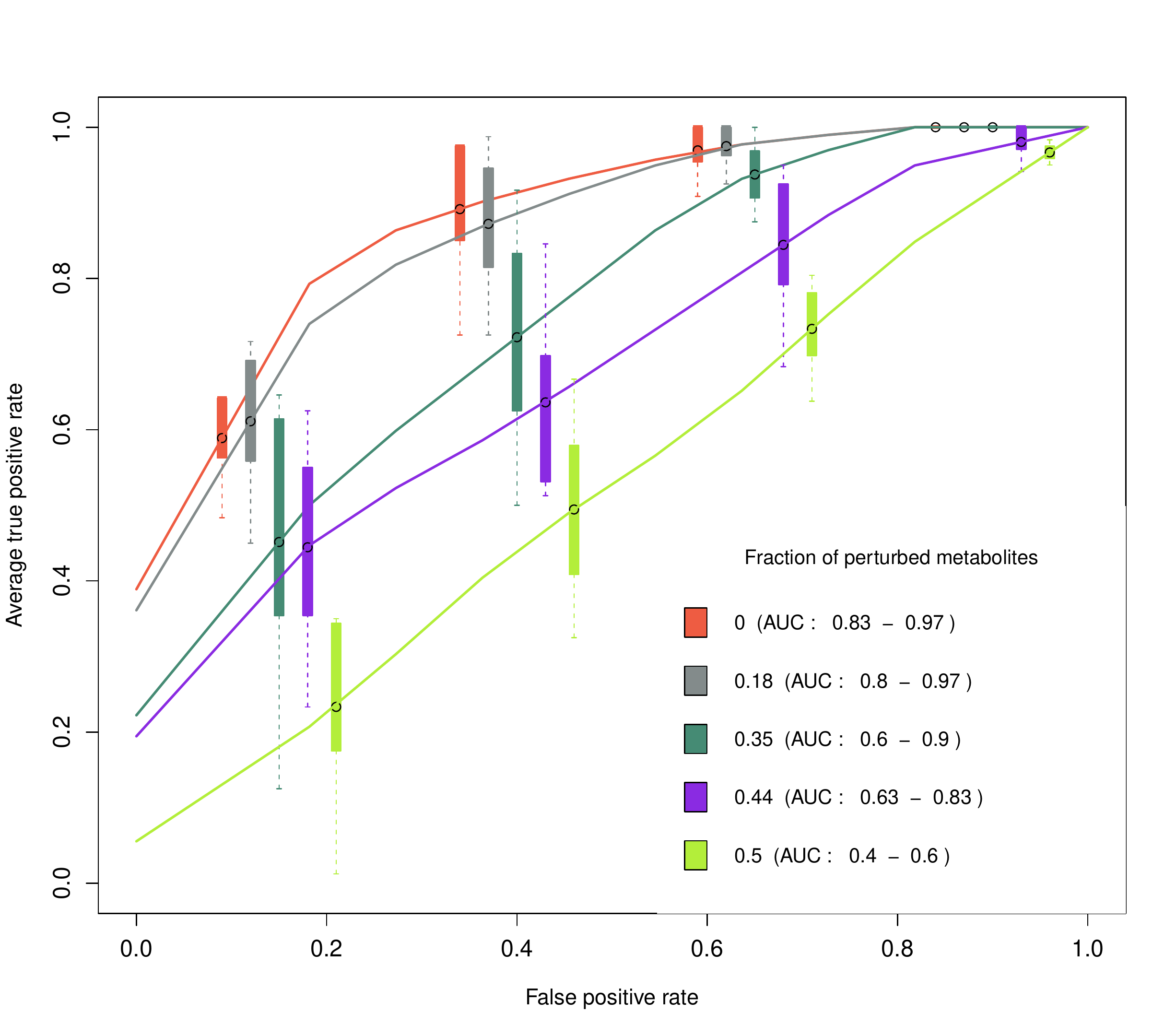}
\caption{Average Receiver Operating Characteristic (ROC) curves for pathway perturbation inference across 10 datasets for different factions of ``falsely" assigned metabolites.}
\label{fig5}
\end{figure}


\subsection{Experimental design}
The covariance structure between metabolites might change drastically as a result of treatment if the latter affects metabolic pathways. The model can be extended to take into account the experimental design. As specified in the previous section, $\alpha_m$ captures the treatment effect for metabolite $m$, $ \gamma_{im}$ represents individual perturbations for metabolite $m$, $\nu_{itm} \vert  \nu_{i,t-1,m}  \sim  N\left(\theta_m \nu_{i,t-1,m}, \sigma_{\nu_m}^2\right)$ represents temporal effects for metabolite $m$ of individual $i$ at time point $t$ in equation~(\ref{eq8}). In addition, we allow covariance structures $\boldsymbol{C} \left( \boldsymbol{\phi}^e \right)$ to be different for the control samples and the cases where $e \in \lbrace \text{cases, controls} \rbrace$ designates experimental groups. This yields the overall hierarchical model:
\begin{eqnarray}
\boldsymbol{x}_{it}^e \vert \boldsymbol{\mu}_{it}, \boldsymbol{C}, \sigma  \sim  N \left(\boldsymbol{\mu}_{it} , \left( \boldsymbol{I}_M-\boldsymbol{C \left( \boldsymbol{\phi}^e\right)} \right)^{-1} \sigma^2 \right) \label{eqx}\\
\mu_{itm} = \alpha_m + \gamma_{im} + \boldsymbol{y}_{it}\boldsymbol{\beta}_m + \nu_{itm} \label{eqmu}\\
\beta_{mk} \vert \lambda_{mk}, \sigma_{\beta_m} \sim N \left( 0, \lambda_{mk}^2 \sigma^2_{\beta_m} \right)\\
\lambda_{mk} \vert \tau \sim \text{St}^+ \left( \tau ,0,1 \right)\\
\gamma_{im} \vert \sigma_{\gamma_m} \sim N\left(0, \sigma_{\gamma_m}^2 \right) \label{eqgam}\\
\nu_{itm} \vert \theta_m, \sigma_{\nu_m} \sim  N\left(\theta_m \nu_{i,t-1,m}, \sigma_{\nu_m}^2\right) \label{eqnu}
\end{eqnarray}
Note that by specifying different dependence parameters for metabolite interactions in cases and controls metabolism, the model is able to identify perturbed pathways by comparing $ \boldsymbol{\phi}^{\text{cases}} $ and $\boldsymbol{\phi}^{\text{controls}}$.

\section{RESULTS}
In this section we perform experiments on both synthetic and real data to investigate whether our algorithm gives reasonable solutions. We first try our method on a simulated dataset in section \ref{secsynt} to get an understanding of the performance of our method. In section \ref{secmet}, we test our method on a data set using metabolomic and bacterial composition in a drug treatment experiment. In the following, we refer to our model as `` iCARH " model for `` integrative CAR Horseshoe " model.

\subsection{Simulation study}
\label{secsynt}
To get better understanding of our method and test its applicability, we first perform our approach on synthetic datasets. We will mainly focus on the ability of our model to infer pathway perturbation.
\subsubsection*{Assessing pathway inference with beta like prior}
In the first simulation our objective is to assess how the beta like prior in equation (\ref{eqphip}) improves the iCARH model. We first fixed the number of pathways P to 11, then simulate the design matrices $\boldsymbol{A_p}$. Specifically, a membership matrix $\boldsymbol{Z}$ with dimensions $M \times P$ is randomly generated based on the density of the number of KEGG pathways in which a single metabolite is involved. Each design matrix $\boldsymbol{A_p}$ is then equal to $\boldsymbol{z_p}\boldsymbol{z_p}^T$ where $\boldsymbol{z_p}$ is the $p$th column of $Z$. Finally, we generated 10 datasets according to the model below in order to assess how our model infers perturbed pathways:
\begin{eqnarray}
\omega \sim \text{Bernoulli} \left( \pi_{\omega} \right)\\
\resizebox{0.95 \linewidth}{!}
{
    $\phi_p^{\text{controls}} \vert \omega \sim \omega N_{[0,1/P\xi_p^2]} \left(\frac{1}{P\xi_p^2}-\rho , \sigma_{\phi}^2 \right) + \left( 1-\omega \right) N \left(0,\psi^2 \right)$
}\\
\resizebox{.95 \linewidth}{!}
{
$\phi_p^{\text{cases}} \vert  \phi_p^{\text{controls}}, \omega \sim  \omega N_{[1/P\xi_p^1,0]} \left(\frac{1}{P\xi_p^1}+\rho , \sigma_{\phi}^2 \right) + \left( 1-\omega \right) \delta_{\phi_p^\text{controls}} $
}
\end{eqnarray} 

with $N_{[a,b]}$ denotes the truncated normal distribution with boundaries $a$, $b$,  $\xi_p^1$ and $\xi_p^2$  are the minimum and maximum eigenvalues of $\boldsymbol{G_p}\boldsymbol{A_p}$. The rest of the parameters is set as follows : number of bacterial variables $K=1$, number of metabolites $M=40$, number of time points $T=7$, number of samples $N=22$, global parameter $\tau$ fixed to~1.2, parameters $\nu_{itm}$, $\gamma_{im}$, $ \mu_{itm}$, $ \boldsymbol{x}_{it}^e$ simulated according to equations~(\ref{eqnu}),~(\ref{eqgam}),~(\ref{eqmu}),~(\ref{eqx}) respectively. 

\begin{figure}[h]
\centering
\includegraphics[scale=0.4]{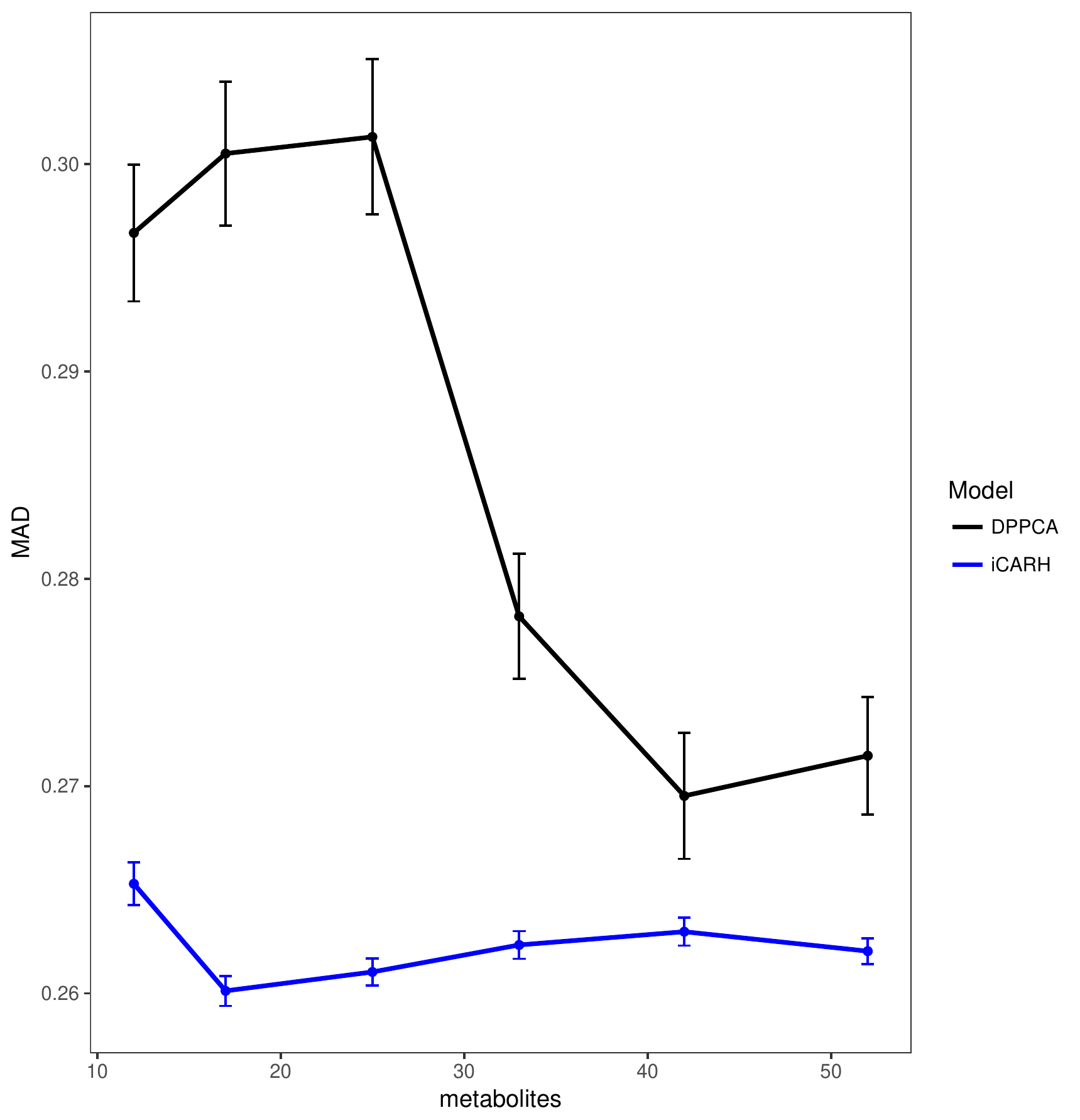}
\caption{Posterior predictive checks for mean absolute deviation (MAD) compared to DPPCA for different numbers of metabolites included. The MAD decreases as the number of metabolites increases. Our model performs clearly better than the DPPCA model.}
\label{fig3}
\end{figure}

We set non-informative uniform priors on $\alpha_m$, $\sigma_{\gamma_{im}}$, $\theta_m$, $\sigma_{\mu_m}^2$. We set an informative prior on $\sigma^2_{\gamma_m} \sim \text{inverse-gamma} \left(1, 0.1 \right)$ as we expect low variability amongst biological samples of the same group. We fix $\pi_{\omega}$ to $0.7$ the proportion of expected perturbed pathways. $\rho$ is fixed to a value of $0.05$, $\sigma_{\phi}^2$ to $0.2$  and $\psi$ to a large value. We compare inference of the model under a uniform prior for $\phi^e_p$ and the prior in equation ~(\ref{eqphip}). Inference is done using 2000 iterations of Hamiltonian Monte Carlo sampling and 1000 warm-up iterations.


The left plot in figure \ref{fig2} shows the boxplots of the Area Under the Curve (AUC) for pathway perturbation inference for 10 simulated datasets with uniform prior on $\boldsymbol{\phi}$ and a beta like prior on $\boldsymbol{\phi}$. We infer perturbations based on the posterior probability that $ \boldsymbol{\phi}^{\text{controls}}$ and $\boldsymbol{\phi}^{\text{cases}}$ are different i.e. the $95\%$ credible interval of $ \boldsymbol{\phi}^{\text{controls}} - \boldsymbol{\phi}^{\text{cases}}$ does not contain zero. The AUC values for the beta like distribution is significantly higher than the AUC values for the uniform distribution. On average pathway peturbation inference under the uniform distribution reduces to a random guess with an average AUC of 0.53. This is likely due to the lack of variance of the uniform distribution. The right plot in figure \ref{fig2} shows the posterior distributions of $ \boldsymbol{\phi}^{\text{cases}} - \boldsymbol{\phi}^{\text{cases}}$ under the uniform prior (white) and the beta like prior (grey) for each pathway. True perturbed pathways are printed in bold on the x axis.

\subsubsection*{Assessing pathway inference against design inaccuracies}
It is very common in metabolomics data to find metabolites that are correlated but not in the same KEGG pathway. In the following simulation we assess how inaccuracies in the covariance structure between metabolites and the design matrices $\boldsymbol{A_p}$ affect the iCARH model. We used the 10 datasets from the previous simulation and perturbed the design matrices by selecting a random fraction of metabolites in each pathway. We then randomly (falsely) assign these metabolites to no pathway, or to different pathways. We similarly run the model for 2000 iterations of Hamiltonian Monte Carlo sampling and 1000 warm-up iterations for each of the fractions $y = 0, 0.18, 0.35, 0.44, 0.5, 0.62$ of perturbed metabolites. Finally, in the same fashion, we assess perturbations based on the $95\%$ credible interval of $ \boldsymbol{\phi}^{\text{controls}} - \boldsymbol{\phi}^{\text{cases}}$. Figure \ref{fig5} is a series of average Receiver Operating Characteristic (ROC) curves across 10 datasets for each of the factions $y$. On average, the performance of our model reduces to a random guess (AUC of 0.5) if $50\%$ of the metabolites in each pathway is perturbed. The AUC of our model reaches 0.97 if no metabolites are perturbed and is about 0.88 if $18\%$ of the metabolites in each pathway are perturbed.

\begin{figure}[!t]
\centering
\includegraphics[scale=0.3]{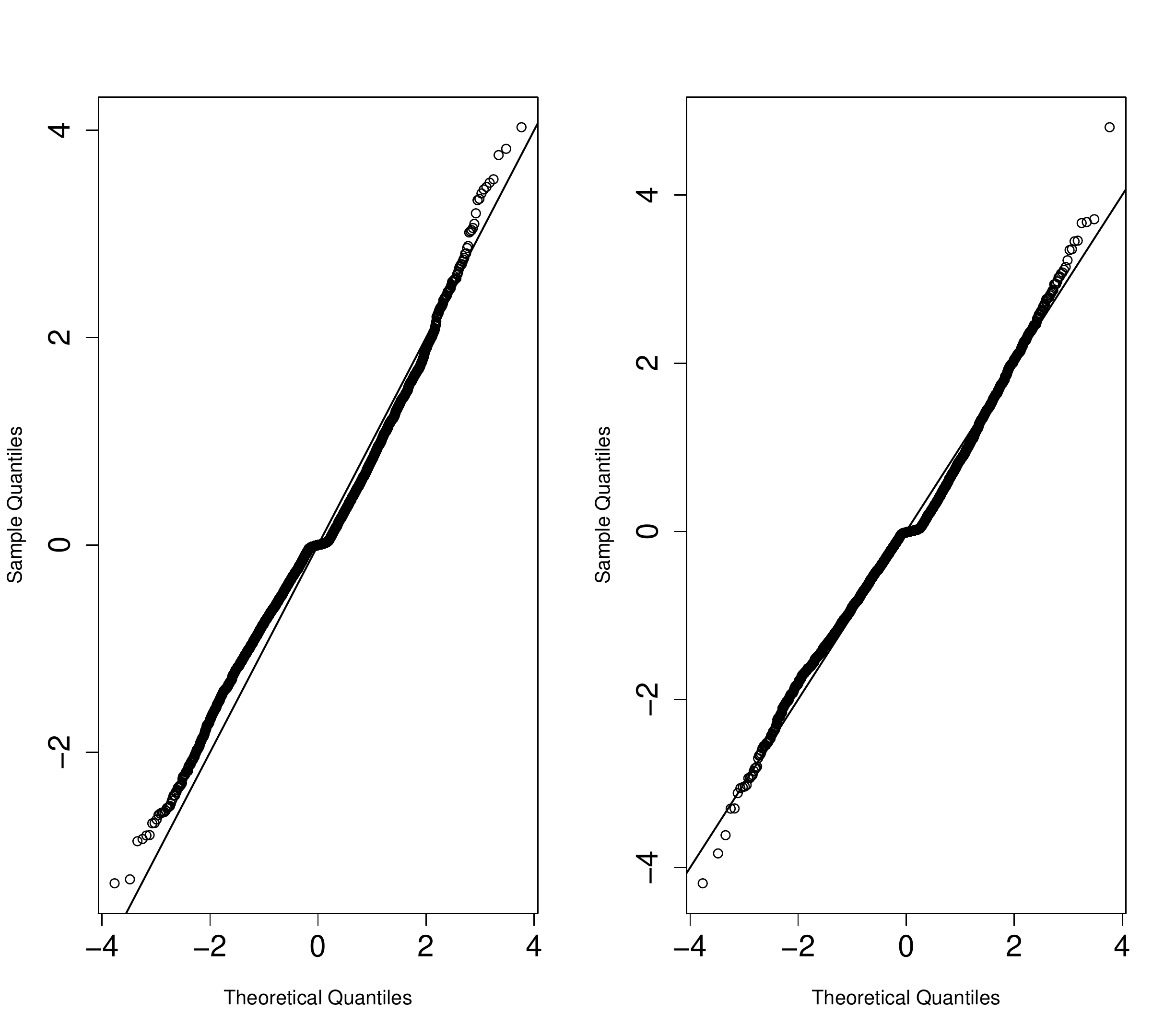}
\caption{Right and left panels show model fit assessment for controls and cases data. Left : quantile-quantile normal plot of $\Psi^{-1}_{\text{cases}} \left( \boldsymbol{x}_{it} - \boldsymbol{\mu}_{it} \right)$. Right : quantile-quantile normal plot of $\Psi^{-1}_{\text{controls}} \left( \boldsymbol{x}_{it} - \boldsymbol{\mu}_{it} \right)$.}
\label{fig7}
\end{figure}

\subsection{Case study}
\label{secmet}

In this section, we test our model on an actual metabolomic data and 16S data for bacterial profiles. In this study we are interested in the influence of metformin on a non-diabetic model. Metformin is the first-line medicine to treat type 2 diabetes. It has also been suggested that metformin has anti-cancer, cardiovascular and anti-aging effects. Because of their very large metabolic capacity, the gut bacteria can influence toxicity and metabolism of drugs. Here, we are particularly looking for metabolic biomarkers indicative of microbiota changes as result of treatment.

The study design is as follows: metabolic profiles of 24 rats are acquired once a week using different mass spectrometry techniques from plasma samples over a period of 9 weeks. Bacterial profiles are acquired using miSeq. The study has allowed for two groups of 12 rats where metformin has been administrated to the second group (weeks 3 to 7) allowing for 2 weeks of acclimatation (weeks 1 and 2) and 2 weeks of recovery (weeks 8 and 9). After data processing and metabolite identification, a total of 56 metabolites and 6 bacteria species are further analysed using our model.  Inference is done using 2000 iterations of Hamiltonian Monte Carlo sampling.

We assess performance of our model for different values of $\tau$ using the Watanabe-Akaike information criterion (WAIC). Tested values of $\tau$ comprise 1, 1.2, 5, 10 with corresponding WAIC values of 7317.296 , 7322.798 , 7317.457 , 7316.476  respectively. WAIC values are very similar for different values of $\tau$ which suggests to use the most selective model with $\tau=10$ as it is the simplest i.e with the smallest number of selected variables.


\begin{figure}[!t]
\includegraphics[scale=0.4]{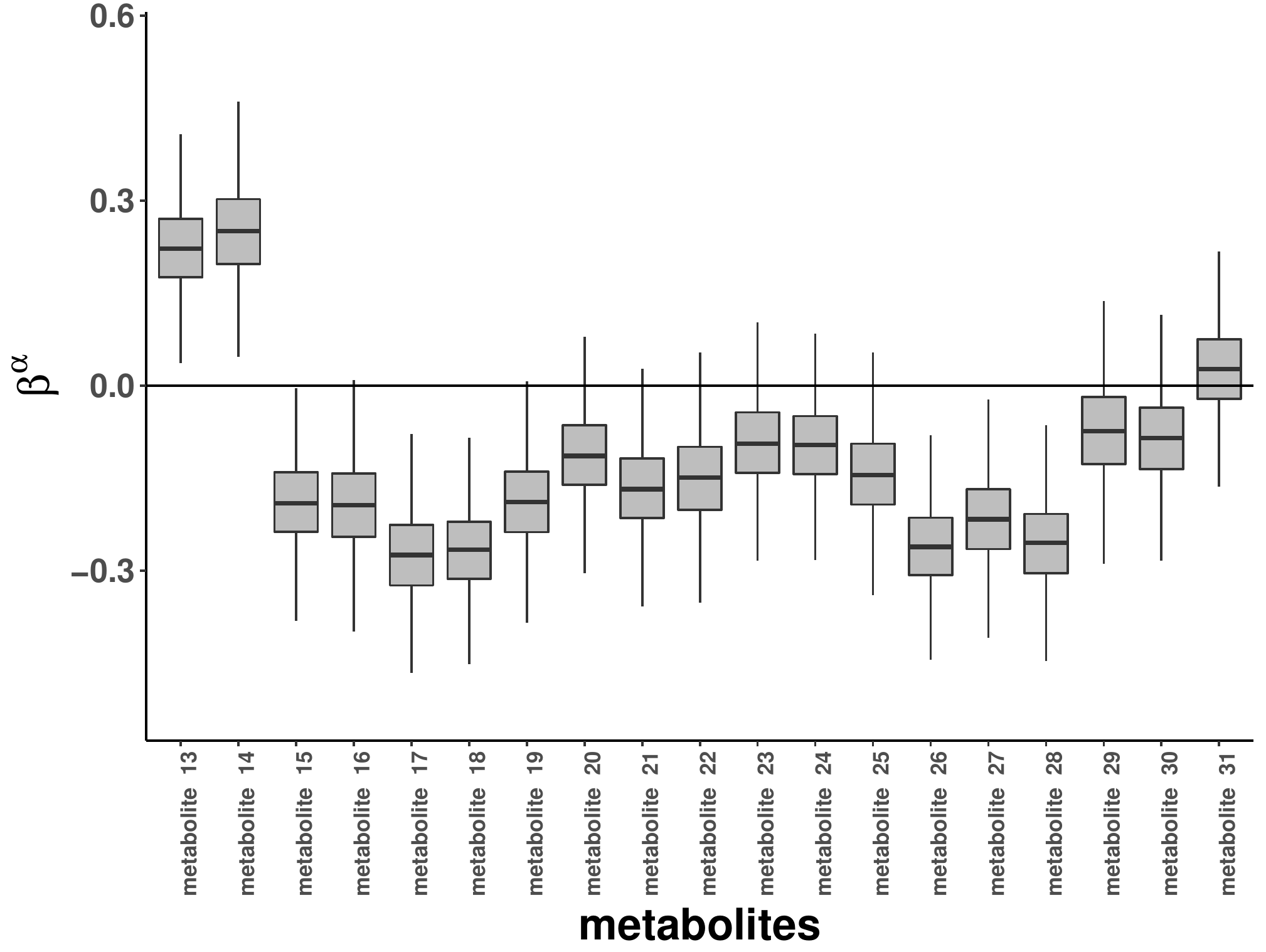}
\caption{Estimates of effects of treatment on metabolite profiles are captured by $\beta^{\alpha}_m$. Only part of the data is plotted as we are only interested in ``metabolite 27". }
\label{fig9}
\end{figure}

%

\subsubsection*{Assessing model fit}
In order to assess our model fit, we perform posterior predictive checks of our model compared to DPPCA \citep{2014dynamic}. The DPPCA model is a multivariate model using PCA, where PCA scores are modeled via a stochastic volatility model. In the Bayesian framework, posterior predictive checks consist in comparing data simulated from the posterior predictive distribution with the observed data. We compared the simulated data and the observed data by means of mean absolute deviations (MADs) between the observed and the simulated covariance matrices for different numbers of metabolites included i.e only part of the data corresponding to these metabolites is considered. The same process was repeated for inference using the DPPCA model \citep{2014dynamic}. Figure \ref{fig3} shows MADs of our model and the DPPCA model. As expected, MADs for both models decrease when the number of metabolites increases. Overall, our model clearly outperforms the DPPCA model. 

In addition to posterior predictive checks, goodness of fit was also checked by using $\Psi^{-1}_e \left( \boldsymbol{x}_{it} - \boldsymbol{\mu}_{it} \right) \sim N\left(0, \boldsymbol{I}_M\right)$ where $\Psi_e$ denotes the Cholesky factor of $\left( \boldsymbol{I}_M-\boldsymbol{C \left( \boldsymbol{\phi}^e\right)}  \right)^{-1} \sigma^2$. Zero-mean and normality were thus checked for $\Psi^{-1}_e \left( \boldsymbol{x}_{it} - \boldsymbol{\mu}_{it} \right)$ (See figure \ref{fig7}).

\subsubsection*{Data results}
Since the administrated drug was also profiled using mass spectrometry, we fit the iCARH model with $\boldsymbol{\alpha}_m = \beta^{\alpha}_m \boldsymbol{y}_{\text{drug}}$. Figure \ref{fig9} is a series of boxplots for $\beta^{\alpha}_m$ for metabolites 13 to 31. We are mainly interested in ``metabolite 27" as it is associated with some bacteria species.

\begin{figure}[b]
\centering
\includegraphics[scale=0.25]{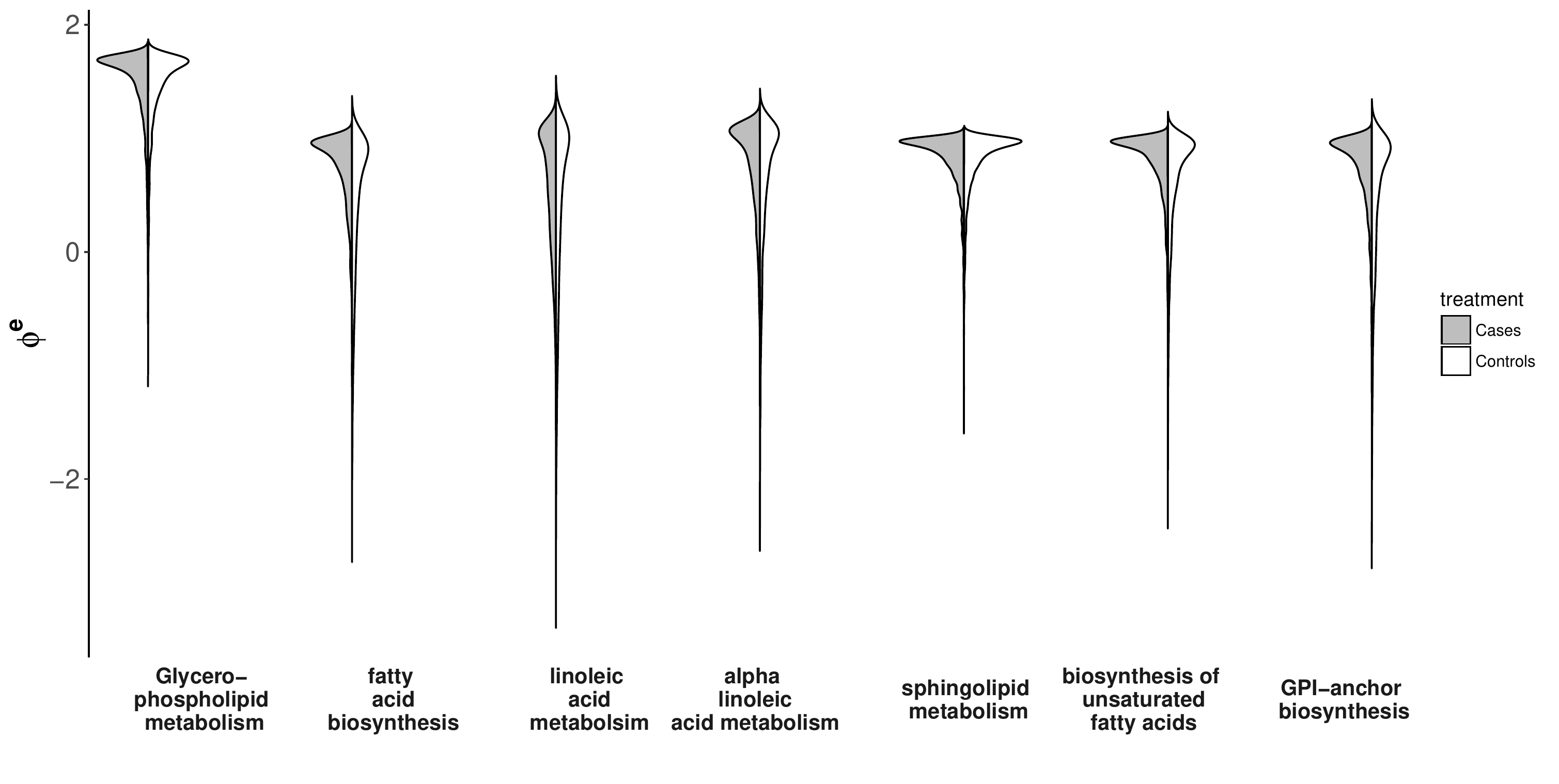}
\caption{Posterior distributions of $\boldsymbol{\phi}^e$ for each pathway for each treatment group. Posterior distributions of $\boldsymbol{\phi}^e$ are very similar for both controls and cases which is indicative of no -or mild- pathway alterations.}
\label{fig8}
\end{figure}

\begin{figure}[!t]
\includegraphics[scale=0.3]{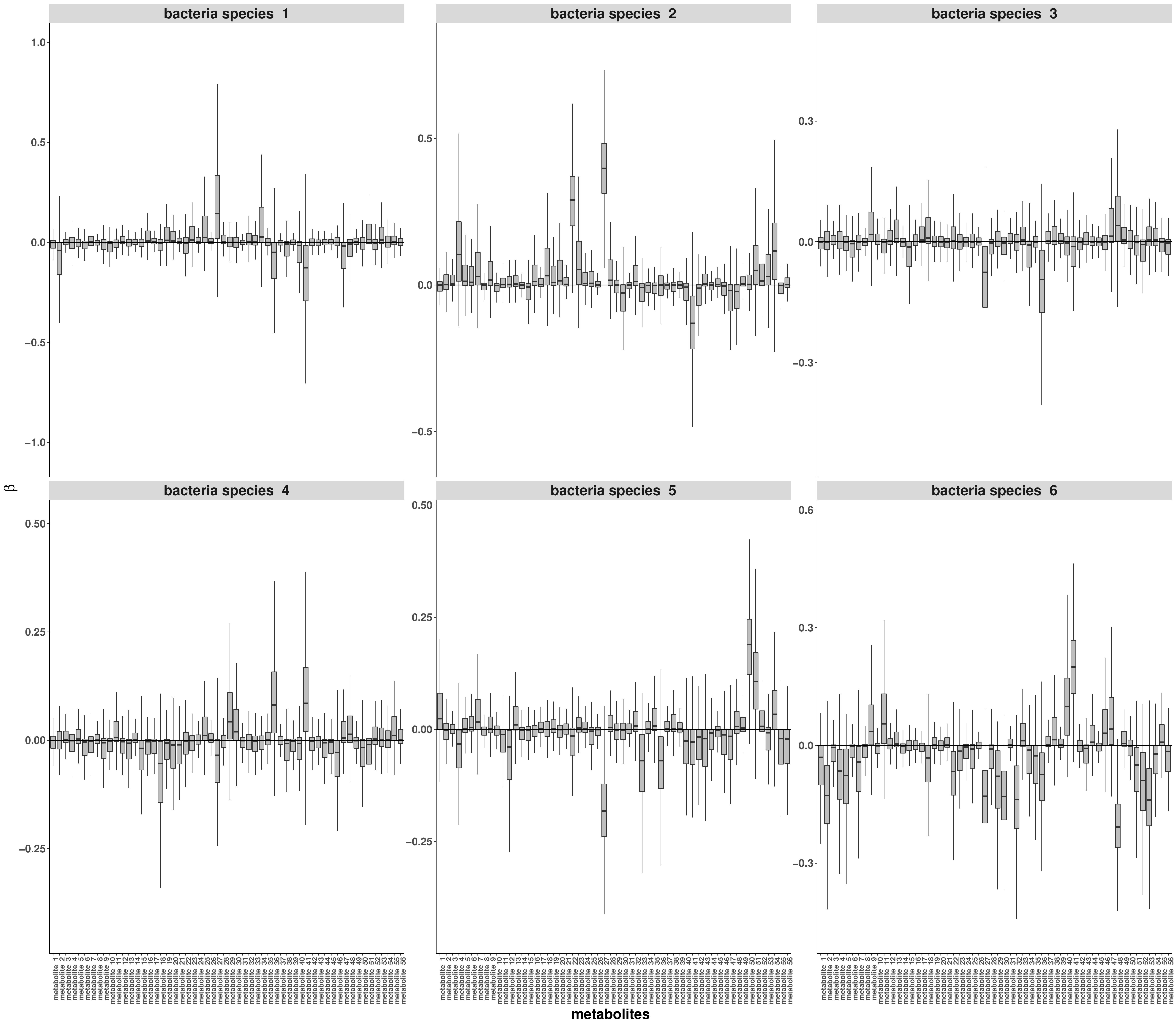}
\caption{Estimates of effects of bacteria on metabolite profiles are captured by $\boldsymbol{\beta}_m$.}
\label{fig4}
\end{figure}

Figures \ref{fig8} and \ref{fig4} show posterior distributions of $\boldsymbol{\phi}^e$ for each pathway and estimates of effects of bacteria on metabolites. Results in section \ref{secsynt} suggest to compare the covariance structure of metabolites in the observed data with the covariance induced by the design matrices in order to have an \textit{a priori} idea on the robustness of pathway inference (See figure \ref{fig5}). For a correlation threshold of $0.3$, about $25\%$ of the metabolites are misspecified in the design matrices which corresponds to an AUC around $0.8$ according to figure \ref{fig5}. If we set a higher correlation threshold, a lower number of metabolites will be misspecified. This supports the use of the iCARH model for pathway perturbation inference for this data.

Estimates of effects of bacteria on metabolite profiles are captured by $\boldsymbol{\beta}_m$. Some metabolites present significant changes along with the bacterial profiles. For example, ``metabolite 27", a hydroxy fatty acid, is associated with alterations in abundance of 4 bacteria species. Figure \ref{fig8} shows that, as a result of treatment, KEGG pathways are not significantly altered. However, distributions of $ \boldsymbol{\phi}^{\text{controls}}$ for ``fatty acids biosynthesis" and ``biosynthesis of unsaturated fatty acids" KEGG pathways are remarkably flatter than the distributions of $ \boldsymbol{\phi}^{\text{cases}}$. These pathways involve the previously identified hydroxy fatty acid metabolite. Our analysis confirms previously reported studies that hydroxy fatty acids might be produced by the gut microbiome \citep{kishino2013, kimura2013}.

\section{DISCUSSION}
Identifying biomarkers in time course metabolic data and inferring significant associations with heterogeneous omic variables is extremely challenging due to the several sources of variations of the data. In addition, existing methods developed to analyze such data are very scarce and have the limitations of  i) overfitting to the few available data points or ii) confounding the experimental and longitudinal variation or iii) ignoring the metabolite interactions or iv) ignoring effects of other omic variables. In this paper, the model we have developed combines several approaches to take into account the different aspects of the data namely the number of time points, the experimental variation captured by $\boldsymbol{\mu}_{it}$, interactions between metabolites captured by $\boldsymbol{\phi}$ and interactions with additional omic variables captured by $\boldsymbol{\beta}_m$. 

Our results demonstrate that our model successfully addresses the main questions of a metabolomic study. Most importantly, our model is able to identify metabolic biomarkers related to treatment, infer perturbed pathways as a result of treatment and find significant associations with additional omic variables. We have shown that providing an informative prior on metabolic pathways and an informative prior over the parameter $\boldsymbol{\phi}$ is a significant improvement over the DPPCA model. Particularly, our model is more robust to slight variations usually observed in short time series data thanks to the small number of covariance parameters needed to estimate compared to DPPCA. We have also shown through simulation that an informative beta like prior compares better than a non-informative uniform prior in inferring significant pathways. On real data, we have investigated how the number of profiled metabolites can affect the predictive ability of the model. 

Several potential extensions arise naturally from our model. In terms of the metabolite interactions component, many research questions can arise. Alternative strategies to modeling metabolite interactions can be examined such as modeling the non-zero elements of the adjacency matrix $\boldsymbol{C}$ of each pathway as random variables. This strategy was adopted in the CAR literature by \cite{lee2013, Rushworth2017} to take into account step changes in spatial variation. Step changes can potentially be useful to model changes in metabolites correlations as a result of treatment. \cite{lee2011} provide an overview of different CAR models used in spatial modeling. The proposed models can be adapted to fit into the metabolomics literature.

From a practical point of view, the model has been fitted using HMC sampling but takes a large amount of time (about 1 hour) mostly because of the variable selection procedure and metabolites interdependence. This could be addressed by using variational Bayes. In fact, variational Bayes inference procedures offer cost-effective inference by means of principled approximations and appealing computational time for high dimensional data. A variational bayes inference of CAR models was proposed by \cite{harrison2010} for high dimensional data, and a variational bayes approach for variable selection was recently proposed by \cite{ormerod2017}.  

\newpage

\newpage\null\thispagestyle{empty}\newpage
\newpage\null\thispagestyle{empty}\newpage

\section{CONCLUSION}
Metabolomics longitudinal profiling techniques are imperative to understand the effect of a drug or a disease across time and can provide enhanced understanding of the underlying biology of the system. In a data integration framework, we have illustrated the use of the CAR model to incorporate metabolites interactions in the model and the horseshoe prior to identify association with heterogeneous omic variables obtained by other omic techniques. The combination of the CAR and horseshoe levels yields the ``integrative CAR Horseshoe" (iCARH) model which we presented in this article.\\
Although, it is computationally expensive, the iCARH model has various appealing features such that it is able to identify metabolic biomarkers related to treatment, infer perturbed pathways as a result of treatment and identify potential associations between heterogeneous omic variables. Clearly, these appealing features open up further research topics.

\subsubsection*{Acknowledgements}

Thanks are due to Panagiotis Vorkas for providing the metabolic and 16S data. Infrastructure support for this work was provided by the NIHR Imperial Biomedical Research Centre.

\subsubsection*{References}

\printbibliography

\end{document}